\documentclass[aps,prd,twocolumn,superscriptaddress,altaffilletter,nobibnotes,nofootinbib,10pt,showpacs,showkeys,preprintnumbers]{revtex4-2}

\usepackage{float}
\usepackage{graphicx}
\usepackage{dcolumn}
\usepackage[dvipsnames]{xcolor}
\usepackage{bm}
\usepackage{lineno}
\usepackage{float}
\usepackage{amsmath}
\usepackage{soul}
\RequirePackage{multirow}

\usepackage{tikz,xcolor,hyperref}

\definecolor{lime}{HTML}{A6CE39}
\DeclareRobustCommand{\orcidicon}{
	\begin{tikzpicture}
	\draw[lime, fill=lime] (0,0) 
	circle [radius=0.16] 
	node[white] {{\fontfamily{qag}\selectfont \tiny ID}};
	\draw[white, fill=white] (-0.0625,0.095) 
	circle [radius=0.007];
	\end{tikzpicture}
	\hspace{-2mm}
}
\foreach \x in {A, ..., Z}{\expandafter\xdef\csname orcid\x\endcsname{\noexpand\href{https://orcid.org/\csname orcidauthor\x\endcsname}
			{\noexpand\orcidicon}}
}

\newcommand{\pte}{$p_{\rm{_T}}$}
\newcommand{\pt}{$p_{\rm{_T}}$~}

\newcommand{\pp}{$pp$~}

\newcommand{\RomanNumeralCaps}[1]
{\MakeUppercase{\romannumeral #1}}

\usepackage{soul}
\usepackage{float}

\definecolor{lime}{HTML}{A6CE39}
\DeclareRobustCommand{\orcidicon}{
	\begin{tikzpicture}
	\draw[lime, fill=lime] (0,0) 
	circle [radius=0.16] 
	node[white] {{\fontfamily{qag}\selectfont \tiny ID}};
	\draw[white, fill=white] (-0.0625,0.095) 
	circle [radius=0.007];
	\end{tikzpicture}
	\hspace{-2mm}
}
\foreach \x in {A, ..., Z}{\expandafter\xdef\csname orcid\x\endcsname{\noexpand\href{https://orcid.org/\csname orcidauthor\x\endcsname}
			{\noexpand\orcidicon}}
}

\begin{document}

\title{The theoretical description of the transverse momentum spectra: a unified model}

\author{Rohit Gupta\orcidA{}}
\affiliation{Department of Physical Sciences, Indian Institute of Science Education and Research (IISER) Mohali, Sector 81 SAS Nagar, Manauli PO 140306 Punjab, India}
\affiliation{Shaheed Mangal Pandey Government Girls Degree College (SMPGGDC), Jananayak Chandrasekhar University (JNCU), Ballia 277001, Uttar Pradesh, India}

\author{Anjaly Menon}
\affiliation{Department of Physics, University of Houston,3507 Cullen Blvd, Houston, Texas 77204-5008, USA}

\author{Shubhangi Jain}%
\affiliation{Department of Physical Sciences, Indian Institute of Science Education and Research (IISER) Mohali, Sector 81 SAS Nagar, Manauli PO 140306 Punjab, India}

\author{Satyajit Jena\orcidB{}}
\email{sjena@iisermohali.ac.in}
\affiliation{Department of Physical Sciences, Indian Institute of Science Education and Research (IISER) Mohali, Sector 81 SAS Nagar, Manauli PO 140306 Punjab, India}

\begin{abstract}
Analysis of transverse momentum distributions is a useful tool to understand the dynamics of relativistic particles produced in high energy collision. Finding a proper distribution function to approximate the spectra is a vastly developing area of research in particle physics. In this work, we have provided a detailed theoretical description of the unified statistical framework in high energy physics. We have tested the applicability of this framework on experimental data by analysing the transverse momentum spectra of pion produced in heavy-ion collision at RHIC and LHC. We have also attempted to explain the transverse momentum spectra of charged hadrons formed in \pp~collision at different energies using the unified statistical framework. This formalism has been proved to nicely explain the spectra of particles produced in soft processes as well hard scattering processes in a consistent manner.
\end{abstract}

\maketitle

\section{\label{sec:intro}INTRODUCTION}
The primary motivation behind the theoretical and experimental studies in particle physics is to enhance our knowledge about the fundamental constituents of matter that make up the universe. One integral component that is elemental in our understanding of the matter content of the universe is the state called Quark-Gluon Plasma (QGP) that was created a few microseconds after the Big-Bang. 
Theoretical calculations based on Lattice Quantum Chromodynamics (LQCD) framework, first predicted the existence of this new state at sufficiently high temperature or baryon densities which were present at the very early stage of Universe expansion. Later, heavy-ion collision experiments performed at Relativistic Heavy Ion Collider (RHIC) \cite{RHIC_exp} and Large Hadron Collider (LHC) \cite{Evans:2008zzb} made it possible to reach energy densities above critical values predicted by Lattice QCD, for the formation of QGP \cite{Jacobs:2004qv,Gyulassy:2003mc,Kolb:2003dz}. 
Since the QGP droplet is being created for an extremely short interval of time ($~10^{-22}$ s) so it is not possible, with present technologies, to directly probe this state. Hence we rely on the kinematics observables such as rapidity, transverse momenta and energy of the final state particles to extract information of such initial state. One such kinematics observable is the transverse momentum \pte-spectra which is the component of momentum in the direction transverse to the beam direction. 

In this paper, we focus on the study of transverse momentum distributions which has proven to be a useful probe for understanding the thermodynamical properties and the evolutionary dynamics of systems produced in relativistic heavy-ion collisions. Our objective is to find a proper distribution function which approximate and explain the transverse momentum spectra. We begin with a short review of developments happened on this path of finding a distribution that can explain particle spectra with better accuracy. Later a solution to the problem is proposed using Pearson distribution \cite{Jena:2020wno}. 

In Section 2, we discuss different statistical frameworks that have been used for understanding thermal QGP systems, including Boltzmann (BG) and Tsallis frameworks. These two models are chosen since these are the most fundamental statistical thermal models, however, a comparison with other models including flow effect etc. is provided in Ref.~\cite{Gupta:2021efj}. We introduce our proposal to the problem using a generalized distribution called Pearson distribution. A detailed formulation of the generalized function is performed,  allowing us to write unified function as an extended form of Tsallis distribution. The validation, goodness of the fit parameter and analyses are given in section 3. Finally, we conclude with a discussion in section 4. 

\section{\label{sec:stat}STATISTICAL APPROACH TO THERMAL QGP SYSTEMS}

Transverse momentum spectra play a pivotal role in enhancing our knowledge of the thermal and bulk properties of QCD matter produced during the heavy-ion collision. However, to extract the parameter of interest, the theoretical model is required that can consistently explain the spectra.  Although QCD is the underlying theory to explain such strongly interacting system, it is a challenging task to apply QCD to explain the spectra in low-\pt regime due to the asymptotic freedom at the perturbative order where the coupling strength is very high. Hence, we resort to statistical thermal approach to explain the spectra. 

Proposal to apply statistical models to explain particle production was given for the first time in 1948 by Koppe \cite{Koppe:1948, Tawfik:2013tza}. Two years later, Fermi introduced a statistical framework \cite{fermi:ptp, Fermi:1951zz} to study the energy distribution of particles coming out from the small volume where large amount of energy is concentrated when two nucleons with high centre of mass energy collide with each other. Although the Fermi model was a reliable description in energy ranges comparable to that of cosmic rays, it breaks down at lower energies. The first systematic description of the mass spectrum of strongly interacting particles based on asymptotic bootstrap principle was formulated by Hagedorn \cite{Hagedorn:1965st, Hagedorn:1967ua} in 1965. The Hagedorn model introduces a limiting temperature $T_0$ which is the highest possible temperature for the strong interaction. Using this model, it was possible to accurately determine the total multiplicity of hadronic particles produced in collisions.Even now, its modified version is used to understand hadronic phase in high energy collisions. These works on statistical description of particle spectra in high energy physics are followed by a large number of papers over the past several decades on the characterization of particle production using statistical mechanics. A detailed review on the application of statistical thermal models can be found in Ref.~\cite{Tawfik:2014eba}. 

In standard thermodynamics, we characterize a macroscopic system using state variables like number density (n), energy density \textbf{$ (\epsilon)$}, pressure (p), temperature (T), chemical potential $(\mu) $ etc. The equilibrium thermodynamic properties of hadronic systems obtained using statistical models can also be characterized by thermodynamic parameters mentioned above. This will finally give insights about the dynamics of the systems in terms of these state variables.  Certain aspects of relativistic kinetic theory are also used in this statistical description. Considering a system of large number of relativistic particles, all the macroscopic quantities required for the thermodynamic description of the system can be derived by using the partition function, which describes the distribution of particles in a thermodynamics system in equilibrium.  The definitions of parameters like energy density, pressure, momentum etc, in terms of the partition function can be found in Ref.~\cite{DeGroot:1980dk}. In the next few subsections, we present the  BG and Tsallis statistical framework which have been used to understand the system of relativistic particles.
\subsection{\label{sec:level2}Boltzmann-Gibbs Statistics}
Considering the system produced in high energy collision to be of thermal origin, most natural choice to describe the distribution of particles will be Maxwell-Boltzmann statistics \cite{Stodolsky:1995ds, Schnedermann:1993ws}. Since the temperature of the system produced in collision is extremely high and Fermi-Dirac as well as Bose-Einstein statistical system tends toward Maxwell-Boltzmann statistics at high temperature, it will be justifiable to use Boltzmann statistics to explain the particle production spectra in the collision.

In general, the expression for the average number of particles in the $s^{th}$ state of a statistical system is given as:
\begin{equation}
{n_s} = \frac{1}{e^{\beta(\epsilon_s- \mu )}\pm 1} 
\end{equation}
If the number of particles in the system is constant, the constraint determining $\mu$ will be given in terms of the following equation.
\begin{equation}
\sum_s{n_s} = \frac{1}{e^{\beta(\epsilon_s- \mu)}\pm 1} = N
\end{equation}
Where, upper and lower sign refers to the case of Bosons and Fermions respectively. When we look at classical limit which is defined by high temperature, the higher energy states will be mostly occupied and the relation $\epsilon_s >> \mu$ will be obeyed.

For keeping $N$ fixed, the $ e^{\beta{\epsilon_s- \mu }} >>1 $ relation must be satisfied. When this is satisfied, the functional form for the number of particles will become exponential like or BG distribution as follows:
\begin{equation}
{n_s} = e^{-\beta(\epsilon_s- \mu )}
\end{equation}
From the standard statistical thermodynamics, we know that the probability of each microstate or the population of particles to occupy each state in a thermal system at equilibrium is an exponential function of energy. 
 For a system of particles following Boltzmann distribution, number density will be given as 
\begin{equation}
n' = \frac{g}{(2\pi)^3} \int d^3p\; exp\left(\frac{\mu -E}{T} \right) 
 \label{eqn:num_density} 
\end{equation}   
which can be written in differential form as
\begin{equation}
\frac{d^3N}{dp^3} = \frac{gV}{(2\pi)^3}\; exp\left(\frac{\mu -E}{T} \right)
\label{eqn:boltz_energy}
\end{equation}
Here, $g$ is the spin degeneracy factor and is equal to $1$ for pseudoscalar mesons (pions, kaons) and $2$ for spin half particles (proton and anti-proton).
Expanding momentum variable in three dimension in polar coordinates will give
\begin{equation*}
d^3p = 2\pi p_T\; dp_T dp_z
\end{equation*} 
where $p_T$ and $p_z$ are transverse and longitudinal momentum respectively.
So, upon equating right hand side of the Eq.~(\ref{eqn:boltz_energy}) we will get
\begin{equation}
E\frac{d^3 N}{dp^3} = \frac{d^2N}{dp_T^2 dy} = \frac{d^2N}{2\pi p_T dp_T dy}
\end{equation}
Here we used the relation $\frac{dp_z}{E}=dy$ where $y$ is the rapidity variable. Using this we will modify Eq.~(\ref{eqn:boltz_energy}) to the form
\begin{equation}
\frac{d^2N}{2\pi p_T dp_T dy}= E \frac{gV}{(2\pi)^3}\; exp\left(\frac{\mu -E}{T} \right)
\end{equation}
We can replace $E$ by $m_T coshy$ where $m_T = \sqrt{m^2 + p_T^2}$ is transverse mass.
Further using the fact that at LHC energies $\mu$ is  vanishing because of approximately equal production of particle and anti-particles and in mid-rapidity region $coshy\simeq 1$ $(y=0)$ we can get
\begin{equation}
\frac{d^2N}{2\pi p_T dp_T dy}=m_T \frac{gV}{(2\pi)^3}\; exp\left(\frac{-m_T}{T} \right)
\label{eqn:BG}
\end{equation}
 The above expression has been used extensively in order to fit the transverse momentum spectra of different particles produced in the collision \cite{Schnedermann:1993ws, Pulawski:2017sct, Basu:2015zra}.

Although this formalism finds its application in many different fields, including high energy physics, there are several issues in explaining the data that needs to be addressed. The application of Maxwell-Boltzmann distribution is limited to the sample where the number of constituents in a system is of the order of Avogadro number $(N_A = 6.023\times10^{23})$. However, in heavy-ion collision, only a few thousand particles are getting produced, limiting the applicability of BG statistics to collision data sample. This difference is also reflected in the deviation of experimental data from the BG function, which fits the experimental data only in a narrow range of \pt and deviate significantly at low as well as high \pte. Further, BG distribution is only applicable to the system where entropy is additive and extensive. However, many physical systems involve long-range interactions and phase space of complex microscopic dynamics that violate BG statistical mechanics and  standard  thermodynamics. Hence a generalization of BG distribution was required to include the non-extensive system.  Tsallis \cite{Tsallis:1987eu} put forward this generalization in 1988, and since then it has been extensively used to study the thermodynamical properties of particle produced in high energy collisions.

\subsection{\label{sec:tsalstat} Tsallis Statistics}

The generalization of Boltzmann-Gibbs theory known as non-extensive statistical mechanics was initially constructed based on an entropy  proposed by Tsallis in 1988. This entropy, called Tsallis entropy has a form that converges to that of BG entropy in a specific limit of its q-parameter. The `q' works as a scaling factor to make standard statistical mechanics applicable to the systems where number of constituents in considerably lower than the Avogadro number. Thus, this parameter in the Tsallis distribution gives the extent of non-extensivity in the thermodynamical system. 

Since the Tsallis statistics has an intrinsic scaling factor in its construction, it is extensively used to explain systems where temperature fluctuations are present around some initial value $T_0$. In such cases, the $q$ parameter, which tells about non-extensivity in the system, can be connected to the variance of temperature \cite{Wilk:1999dr,Wilk:2012zn} as:
\begin{equation}
q - 1 = \frac{Var(T)}{\left\langle T\right\rangle^2 }
\end{equation} 
One major modification in the algebra related to Tsallis statistics is the introduction of q-exponential and q-logarithm given as:
\begin{equation}
exp_q(x) = \left[ 1 - (q-1)x\right] ^{-\frac{1}{q-1}}
\end{equation}\label{eqn:q-exp}
and
\[
    ln_q(p_i) = 
\begin{cases}
    ln(p_i) ,& \text{if } p_i\geq 0,q= 1\\
    \frac{p_i^{1-q}-1}{1-q},& \text{if } p_i\geq 0,q\neq 1\\
    undefined,              & \text{if } p_i\leq 0
\end{cases}
\]
Non-extensive entropy as proposed by Tsallis \cite{Tsallis:1987eu} is defined as:
\begin{align}
    S_q &= -k \sum_{i}p_i^q ln_{q}(p_i) \\
        &= -k \sum_{i}p_i^q \frac{p_i^{1-q}-1}{1-q}\\ 
        &= -k \sum_{i} \frac{p_{i} - p_i^q }{1-q} \\ 
        &=  k \frac{1-\sum_{i}p_i^q}{q-1}
\end{align}
which in the limit $q\longrightarrow1$ gives standard Gibbs entropy
\begin{equation}
S = -k \sum_{i} p_i ln(p_i)
\end{equation}
%
As we discussed earlier, the BG approximation to transverse momentum spectra fails at lower and higher momentum ranges.In contrast, Tsallis approximation works better, and numerous studies have been performed using the same motivation of Tsallis statistics to particle production spectra \cite{Cleymans:2012ya,Wilk:2000yv,Wilk:1999dr}. The thermodynamical aspects of this formalism and its foundations and applications are discussed in \cite{tsallisbook,tsallisgellmannbook}. In this context, we can easily obtain Tsallis statistical distribution used for fitting to transverse momentum data from BG distribution by replacing the exponential in BG function by q-exponential.
And the distribution function used for fitting to particle spectra  can be derived accordingly where, $m_T$, \pte, $T$, $g$, and $V$ have the same meaning as in BG distribution and $y$ is the rapidity variable.
\begin{equation}\label{TSALLIS}
\frac{1}{2\pi p_T} \frac{d^2 N}{dp_T dy} = \frac{gV m_T}{(2 \pi)^3} \left[1+(q-1)\frac{m_T - \mu}{T}\right]^{-\frac{q}{q-1}}
\end{equation}
The definition of Tsallis statistical version of Fermi-Dirac and Bose-Einstein distributions along with corresponding entropy functionals are given in \cite{Cleymans:2012ya}. As we can define number density, energy density, pressure etc in BG formalism \cite{DeGroot:1980dk}, similar can be done using non-extensive relativistic kinetic theory. The only difference is that the Tsallis distribution function will be raised to power $q$. 
In standard thermodynamics, we are familiar with the constraints on the total number of particles, $N$ and energy, $E$ in the system. Given the distribution function $f_i$,
\begin{equation}
N = \sum_i f_i
\end{equation}
\begin{equation}
E = \sum_i f_iE_i
\end{equation}
 whereas in Tsallis case, above constraints must be redefined in following way with function raised to a power of $q$.
\begin{equation}\label{constraint}
  \begin{aligned}
N &= \sum_i f_i^q\\
E &= \sum_i f_i^{q}E_i
\end{aligned}
\end{equation} 
 The appropriate definition of entropy  to accommodate  positive entropy production according to second law of thermodynamics is given in \cite{Biro:2011bq}. Further in classical limit, Tsallis entropy will have following functional form as explain in \cite{Cleymans:2012ya}. 
\begin{equation}
    S_T = -g\sum_i (f_{i}^{q}ln_{q}f_i - f_i) 
\end{equation}
Here, $ln_q(x)$ is the q-logarithm and is defined as:
\begin{equation}
ln_q(x) \equiv \frac{x^{1-q} - 1}{1-q}
\end{equation}
So, on expanding the form of entropy we will get
\begin{equation}
S_T = g \sum_i \left[  \frac{q f_i}{q-1} - \frac{f_i^q}{q-1}\right] 
\end{equation}
By maximising above entropy under the constraints given in Eq.~(\ref{constraint}) we will get variational equation:
\begin{equation}
\begin{aligned}
\frac{\delta}{\delta f_i}\left[ S_T + \alpha\left( N - \sum\limits_{i} f_i^q\right) + \beta\left( E - \sum\limits_{i} f_i^q E_i\right)  \right] = 0
\end{aligned}
\label{eq:lagrange}
\end{equation}
Here $\alpha$ and $\beta$ are the Lagrange multipliers for total number of particles and total energy respectively. On solving this equation we will get distribution function $(f_i)$. 
\begin{equation}
f_i = \left[ 1 + (q-1)\frac{E_i - \mu}{T}\right]^{-\frac{1}{q-1}} 
\label{eq:tsallis_dist}
\end{equation}
It can be easily shown that the entropy defined above will give Tsallis distribution function under its extremization. However, this distribution function must be thermodynamically consistent, which can be shown by checking whether the relations among thermodynamic parameters are indeed obeyed. In Ref.~\cite{Cleymans:2012ya}, the Tsallis function has been proved to be consistent with respect to the laws of standard thermodynamics.
\subsection{Hard processes and limitation in Tsallis statistics}
Tsallis statistics has been used extensively to describe high energy systems, particularly for describing transverse momentum spectra. In Ref.~\cite{Cleymans:2012ya}, after proving thermodynamic consistency of Tsallis distribution function, a modified form is proposed and the fit details are also presented. In another work \cite{Wilk:2000yv,Wilk:1999dr}, a review on implementing non-extensive statistical mechanics for the description of heavy-ion collisions is given, together with new interpretations for the non-extensivity parameter $q$. In \cite{Wilk:2008ue} and \cite{Biro:2005uv}, different power laws used in explaining high energy processes and the importance of non-extensive formalism is discussed. Another work from literature is \cite{Si:2017cyg}, where \pte-spectra of negatively charged pions are fitted with a standard function and its Tsallis form and a comparison is made. 
The non-extensive Tsallis approach provides better fits and, explains heavy-ion collisions more appropriately as compared to the standard Boltzmann and power-law approaches. The plots of \pte-spectra of $ \pi^{+}$ and $ \pi^{-}$ particles fitted to Tsallis function is displayed in \cite{De:2007zza}, where satisfying agreement between the experimental data and function is established.
However, it is known that the exponential function or standard BG theory can only take care of soft \pt region of the hadronic spectra where the particles produced will have small transverse momenta. Whereas QCD calculations have shown that power-law functions can explain spectra of particles produced in hard scattering processes \cite{Si:2017cyg, Arnison:1982ed}. 

For low-\pt part of the spectra (corresponding to ``soft sector"), methods described above and their variants have been used extensively to study the spectra. However, particle production in high \pt regime is dominated by hard QCD scattering processes, and corresponding spectra follow a power law. It is difficult to explain these two regions of the spectra using a single probability distribution function. Hence, several two-component models have been devised to explain the full range of the spectra. However, it is difficult to determine the clear boundary between the two regions of the spectra \cite{Tawfik:2019wze}. Notably, we intend to find a master distribution function which describes the whole region of \pte-spectra (both soft and hard part) in a unified manner.

Although the Tsallis approach takes care of the generalization upto some extent \cite{De:2007zza}, it has been shown in a recent paper \cite{Si:2017cyg, Tawfik:2019wze} that the Tsallis distribution, in its purest form, can describe only the low-\pt range of the spectra, which belongs to the particles produced in soft excitation process. In  Ref.~\cite{Azmi:2015xqa, Cirto:2014sra},  compatibility of Tsallis statistics with \pte-spectra at large transverse momenta has been established in $pp$ collision, however, apparent deviation from the data is observed in the higher momentum range of the spectra of particles produced in the heavy-ion collision as can be seen in the figures. Some efforts are being made to modify Tsallis statistics to explain the hard part of the spectra \cite{Wong:2013sca, Wong:2014uda}, however, search for a consistent framework to explain the spectra is still an open question.

Since perturbative QCD can be used to describe hard scattering processes, it is possible to extract the form of \pte-spectra at these regions. And, the calculations suggest that the spectra will have the form of inverse power law which is expressed as
\cite{Si:2017cyg,Arnison:1982ed,Wong:2012zr,Biyajima:2016fpb,Michael:1976pz,Michael:1977hx}: 
\begin{equation}
f(p_T) = \frac{1}{N} \frac{dN}{dp_T} = A p_T\left(1 + \frac{p_T}{p_0} \right)^{-n}  
\label{eq:hard}
\end{equation}
Where $\textit{p}_0$ and $n$ are fitting parameters and $A$ is the normalization constant related to free parameters. This QCD inspired model was proposed by Hagedorn \cite{Hagedorn:1983wk} to describe data of invariant cross section of hadrons as a function of \pte. Our proposal to the question above stated is to combine inverse power-law and Tsallis distribution using Pearson distribution. This statistical approach is discussed in detail in the following sections.
\subsection{\label{sec:perstat}A generalization of Tsallis statistics}
A comprehensive approach is to look into the parent equation of Tsallis family of equations, which must have more parameters yet controllable mechanism to hold thermodynamics laws. Following the same argument, we look upon the Pearson distribution, which intrinsically deals with more parameters and controlled by the first four fours moments of the distribution, making a perfect choice for our case.
The proposal for Pearson distribution was first given by Karl Pearson in 1895 \cite{Pearson343} and subsequently modified in 1901 and 1916. His main idea was to categorize any distribution function based on the first four moments related to mean, standard deviation, skewness and kurtosis of the distribution. Moments are defined for specifying the shape of any probability distribution. The first moment or the mean locates the centre of the distribution, whereas variance gives the spread or dispersion in the data about the mean. Other two are called shape parameters, among which skewness provides the degree of asymmetry in the distribution around the mean and kurtosis specifies the relative peakedness or flatness of the distribution. Characterization of any statistical data involves the specification of skewness and kurtosis.

Gaussian, Beta, Gamma, inverse-gamma, exponential, Student's T-distribution are all special cases in Pearson distribution and belong to Pearson family of curve. Thus, it is considered as the most general distribution and has been used in many different fields like geophysics, bio-statistics and financial marketing. 
It is a family of continuous probability distributions, whose densities $p(x)$ satisfy the following differential equation \cite{pollard}.  
\begin{equation}\label{diffe}
 \frac{1}{p(x)}\frac{dp(x)}{dx} + \frac{a + x}{b_0 + b_1 x + b_2 x^2} = 0
\end{equation}
where the parameters $a, b_0, b_1, b_2$ can be related to first four central moments as follows:
\begin{equation}
a = b_1 = \frac{m_3(m_4 + 3 m_2^2)}{10m_2m_4 - 18 m_2^3 - 12 m_3^2}
\end{equation}
\begin{equation}
b_0 = \frac{m_2(4m_2m_4 - 3 m_3^2)}{10m_2m_4 - 18 m_2^3 - 12 m_3^2}
\end{equation}
\begin{equation}
b_2 = \frac{2m_2m_4 - 6 m_2^3 - 3 m_3^2}{10m_2m_4 - 18 m_2^3 - 12 m_3^2}
\end{equation}
Here, $m_1, m_2, m_3$ and $m_4$ are the first four central moments with $m_1 =0$. Pearson curves are classified into 12 different types based on the root of the quadratic equation in the denominator of differential equation. Therefore, Pearson criteria which will decide the type of distribution is the sign of discriminant of the quadratic equation which is expressed as,
\begin{equation}
k = \frac{b_1^2}{4b_0 b_2}
\end{equation}
A table including different types of Pearson distribution along with Pearson criteria and condition on parameters can be found in Ref.~\cite{halnpg, Behera:2017xwg}. Further, solving the differential equation (\ref{diffe}) we can find the Pearson density as follows:
\begin{align}\label{eq:variable}
    p(x) &= C^{'} \exp { \int -\frac{P(x)}{Q(x)} dx }\\ 
         &= C^{'} \exp { \int -\frac{a_{0}+a_{1}x}{b_0 + b_{1}x + b_{2}x^{2}} dx }
\end{align}
We can express the quadratic equation in following form,  
\begin{equation}
    b_0 + b_{1}x + b_{2}x^{2} = b_2(x+\alpha)(x+\beta)
\end{equation}

\begin{align}
    p(x) &= \frac{C'}{-b_2} \exp { \int \frac{a_{0}+a_{1}x}{(x+\alpha)(x+\beta)} dx }\\
         &= C \exp{ \int \left( \frac{v}{x+\alpha} + \frac{w}{x+\beta} \right) dx}
\end{align}
Where $v$ and $w$ have following definition:
\begin{align}
 v = -\frac{a_{0}-a_{1} \alpha}{\alpha - \beta} && w = \frac{a_{0}-a_{1} \beta}{\alpha - \beta} 
\end{align}
After integration,
\begin{align}
    p(x) &= C \exp \left\lbrace ln(x+\alpha)^{v} + ln(x+\beta)^{w} \right\rbrace \\
         &= C (x+\alpha)^{v} (x+\beta)^{w}
\end{align}
A general solution can be written as in Eq.~(\ref{eq:solution}) where $C$ is a normalization constant and $e, f, g$ and  $h$ are free parameters.
\begin{equation}
p(x) = C(e + x)^f (g +x)^h 
\label{eq:solution}
\end{equation}
Pearson family of distributions give an advantage of extra free parameter yet preserving all previous models.  At this stage going back to our initial argument, we can show how the Pearson equation is reducible to Tsallis in limiting case. The Pearson function can be expressed as an extended version of Tsallis distribution. It is easy to see that the Pearson distribution converges to exponential when the numerator, $P(x)$ and denominator, $Q(x)$ in Eq.~(\ref{eq:variable}) becomes constant and unity respectively. Similarly, we can derive the limit of Pearson parameters at which it will reduce to Normal or Gaussian distribution. For this, $P(x)$ has to be of linear form and $Q(x)$ has to be unity. Since Pearson density reduces to exponential at some limit, it is possible to find a relation between Tsallis, which is a generalized Boltzmann, and the Pearson function.
The Eq.~(\ref{eq:solution}) can be rewritten in following form by doing simple algebra.
\begin{equation}\label{PEARSON}
p(x) = B \left( 1+ \frac{x}{e}\right)^f  \left( 1 + \frac{x}{g}\right)^h   
\end{equation}
Up to some normalization constant $B = C e^f g^h$. Now if we replace $g = \frac{T}{q-1}$, $h = -\frac{q}{q-1}$, $f = -n$ and $e = p_0$ we will get:
\begin{equation}
p(x) =  B  \left( 1 + \frac{p_T}{p_0}\right)^{-n} \left( 1 + (q-1)\frac{p_T}{T}\right)^{-\frac{q}{q-1}}
\end{equation}
where, 
\begin{equation}
B = C \frac{1}{(p_0)^n} \left(\frac{T}{q-1} \right)^{-\frac{q}{q-1}}
\end{equation}
 This makes it a perfect choice to fit the particle spectra with this function.
\begin{equation}\label{pearson_final}
    \frac{1}{2\pi p_T} \frac{d^2 N}{dp_T dy} = B'  \left( 1 + \frac{p_T}{p_0}\right)^{-n} \left( 1 + (q-1)\frac{p_T}{T}\right)^{-\frac{q}{q-1}} 
\end{equation}
where $B'=B\times \frac{V}{(2\pi)^3}$ with the additional $\frac{V}{(2\pi)^3}$ comes when we move from summation to integration.
Hence, it is inferred that the unified distribution is a generalized form of Tsallis distribution and can be shown to have two parts. In reference to Eq.~(\ref{eq:hard}), the inverse power law term in the above equation can be considered as the hard scattering part in the extended Tsallis form of distribution.

At the same time, the backward compatibility of the distribution is making it more prominent and stable equation to be considered while proposing a generalization. In the context of unified distribution, since it is proposed to be a generalization of Tsallis hence, it should reduce to Tsallis distribution under some limit on parameters. We observed that the unified distribution (\ref{pearson_final}) is backward compatibility in the limit $n = -1$ and $p_0 = 0$. Which means that in the limit discussed above, the unified distribution reduces to the Tsallis distribution preserving all thermodynamics properties. Hence, we can describe unified distribution as the generalization of Tsallis distribution with an additional part to explain higher-\pt part of the spectra corresponding to the particles produced in hard scattering processes.
\subsection{Thermodynamical consistency check for unified distribution}
The proposed equation must pass the thermal test, in this context, a procedure similar to that used for Tsallis framework is followed with a modified form of Tsallis entropy. To show the thermodynamical consistency of the distribution, following relation \cite{Cleymans:2012ya} must be satisfied:
\begin{equation}\label{eq:rel1}
T = \frac{\partial \epsilon}{\partial s}\biggr\vert_n 
\end{equation}
\begin{equation}\label{eq:rel2}
\mu = \frac{\partial \epsilon}{\partial n'}\biggr\vert_s
\end{equation}
\begin{equation}\label{eq:rel3}
n' = \frac{\partial P}{\partial \mu}\biggr\vert_T
\end{equation}
\begin{equation}\label{eq:rel4}
s = \frac{\partial P}{\partial T}\biggr\vert_\mu
\end{equation}
 Constraint equation of total number of particles and total energy remains same as in Tsallis distribution:
\begin{equation}
  \begin{aligned}
N &= \sum_i f_i^q\\
E &= \sum_i f_i^{q}E_i
\end{aligned}
\end{equation} 
In case of unified distribution
\begin{equation}
E \frac{d^3 N}{dp^3} = B'  \left( 1 + \frac{E}{p_0}\right) ^{-n} \left(  1 + (q-1)\frac{(E-\mu)}{T}\right)^{-\frac{q}{q-1}}
\end{equation}
\begin{equation}
\frac{d^3 N}{dp^3} = \frac{B'}{E} \left( 1 + \frac{E}{p_0}\right) ^{-n} \left(  1 + (q-1)\frac{(E-\mu)}{T}\right)^{-\frac{q}{q-1}}
\end{equation}
We can simplify above equation to
\begin{equation}
\frac{d^3 N}{dp^3} = B' f_E f_{Ta}^q
\end{equation}
where
\begin{equation}
f_E = \frac{1}{E} \left( 1 + \frac{E}{p_0}\right)^{-n} 
\end{equation}
\begin{equation}
f_{Ta} = \left(  1 + (q-1)\frac{(E-\mu)}{T}\right) ^\frac{-1}{q-1}
\end{equation}
Hence, we have
\begin{equation}
\frac{d^3 N}{dp^3} = \frac{V}{(2\pi)^3}\left\lbrace \left( B f_E\right)^{\frac{1}{q}}f_{Ta} \right\rbrace^q 
\end{equation}
or more generally,
\begin{equation}
\frac{d^3 N}{dp^3} = \frac{V}{(2\pi)^3}f_i^q
\end{equation}
where
\begin{equation}
f_i =  \left(B  f_{E_i}\right)^{\frac{1}{q}}f_{Ta_i}
\label{eq:pear_dist}
\end{equation}
Entropy in case of unified distribution is given as
\begin{equation}
S_p =\sum\limits_{i}\left[ \frac{q f_i}{(q-1)\left(B f_{E_i}\right)^{\frac{1}{q} - 1} } - \frac{f_i^q}{q-1} \right]
\label{eq:pear_entropy}
\end{equation}
With this form of entropy, we can solve Eq.~(\ref{eq:lagrange}) to get the form of distribution function Eq.~(\ref{eq:pear_dist}).  

For consistency check, we have to prove the basic thermodynamic relations given in equations (\ref{eq:rel1}), (\ref{eq:rel2}), (\ref{eq:rel3}) and (\ref{eq:rel4}).

\subsubsection{Relation 1}
First of the relations above is the derivative of energy density with respect to entropy density given as:
\begin{equation}
T = \frac{\partial \epsilon}{\partial s}\biggr\vert_{n'} 
\end{equation}
Solving the right hand part of the equation
\begin{equation}
     \frac{\partial E}{\partial S}\biggr\vert_{n'} = \frac{\frac{\partial E}{\partial T}dT+{\frac{\partial E}{\partial \mu}d \mu}}{\frac{\partial S}{\partial T}dT+{\frac{\partial S}{\partial \mu}d \mu}}
\end{equation}
\begin{equation} \label{eq:rel1_initial}
     \frac{\partial E}{\partial S}\biggr\vert_{n'} = \frac{\frac{\partial E}{\partial T}+{\frac{\partial E}{\partial \mu}\frac{d \mu}{dT} }}{\frac{\partial S}{\partial T}+{\frac{\partial S}{\partial \mu}\frac{d \mu}{dT}}}
\end{equation}
    In this relation, $n'$ is constant which add additional constraint
\begin{equation}
dn' = \frac{\partial n'}{\partial T}dT +  \frac{\partial n'}{\partial \mu}d\mu = 0
\end{equation}
\begin{equation}
    \frac{d \mu}{d T} = -\frac{\frac{\partial n'}{\partial T}}{\frac{\partial n'}{\partial \mu}}
\end{equation}
Solving for components of Eq.~(\ref{eq:rel1_initial})
\begin{equation}
    \frac{\partial E}{\partial T} = \sum\limits_{i} qf_i^{q-1}E_i\frac{\partial f_i}{\partial T}
\end{equation}
\begin{equation}
     \frac{\partial E}{\partial \mu} = \sum\limits_{i} qf_i^{q-1}E_i\frac{\partial f_i}{\partial \mu}
\end{equation}
\begin{equation}
    \frac{\partial S}{\partial T} =\sum\limits_{i} \frac{q}{(q-1)\left(B f_{E_i}\right)^{\frac{1}{q} - 1}}\frac{\partial f_i}{\partial T} - \frac{qf_i^{q-1}}{q-1} \frac{\partial f_i}{\partial T}
\end{equation}
\begin{equation}
    \frac{\partial S}{\partial \mu} =\sum\limits_{i} \frac{q}{(q-1)\left(B f_{E_i}\right)^{\frac{1}{q} - 1}}\frac{\partial f_i}{\partial \mu} - \frac{qf_i^{q-1}}{q-1} \frac{\partial f_i}{\partial \mu}
\end{equation}
\begin{equation}
    \frac{\partial n'}{\partial T} =\frac{1}{V} \sum\limits_{i} qf_i^{q-1}\frac{\partial f_i}{\partial T}
\end{equation}
\begin{equation}
     \frac{\partial n'}{\partial \mu} =\frac{1}{V} \sum\limits_{i} qf_i^{q-1}\frac{\partial f_i}{\partial \mu}
\end{equation}
Simplifying the numerator of Eq.~(\ref{eq:rel1_initial}) 
\begin{equation}
    \frac{\partial E}{\partial T}+{\frac{\partial E}{\partial \mu}\frac{d \mu}{dT}} = \sum\limits_{i} qE_if_i^{q-1} \frac{\partial f_i}{\partial T} - \frac{\sum\limits_{i,j}q^2E_i (f_if_j)^{q-1}\frac{\partial f_i}{\partial \mu}\frac{\partial f_j}{\partial T}}{\sum\limits_{j}qf_j^{q-1}\frac{\partial f_j}{\partial \mu}}
\end{equation}
This can be further reduced to 
\begin{equation}\label{eq:rel1_part1}
   \frac{\partial E}{\partial T}+{\frac{\partial E}{\partial \mu}\frac{d \mu}{dT}} =   \frac{\sum\limits_{i,j}qE_i(f_if_j)^{q-1}C_{ij}}{\sum\limits_{j}f_j^{q-1}\frac{\partial f_j}{\partial \mu}}
\end{equation}
where
 \begin{equation}
     C_{ij} =\left\{ \frac{ \partial f_i}{\partial T}\frac{\partial f_j}{\partial \mu} - \frac{ \partial f_i}{\partial \mu}\frac{\partial f_j}{\partial T} \right\}
 \end{equation}
 Similarly solving for the denominator part of Eq.~(\ref{eq:rel1_initial})
 \begin{equation}\label{eq:rel1_part2}
   \frac{\partial S}{\partial T}+{\frac{\partial S}{\partial \mu}\frac{d \mu}{dT}} = \frac{\sum\limits_{i,j}\left\{\frac{qf_j^{q-1}}{(q-1)\left(B f_E\right)^{\frac{1}{q} - 1}} -  \frac{q(f_if_j)^{q-1}}{q-1}\right\}C_{ij}}{\sum\limits_{j} f_j^{q-1}\frac{\partial f_j}{\partial \mu}} 
\end{equation}
From Eq.~(\ref{eq:rel1_part1}) and Eq.~(\ref{eq:rel1_part2}) we get
\begin{equation}
    \frac{\partial E}{\partial S}\biggr\vert_{n'} = \frac{\sum\limits_{i,j} qE_i (f_if_j)^{q-1}C_{ij}}{\sum\limits_{i,j} \left(\frac{q}{q-1}\right) \left[\frac{f_j^{q-1}}{(Bf_{E_i})^{\frac{1}{q}-1}}- (f_if_j)^{q-1}\right]C_{ij}}
\end{equation}
\begin{equation}
     \frac{\partial E}{\partial S}\biggr\vert_{n'} = \frac{T \sum\limits_{i,j} E_i (f_if_j)^{q-1}C_{ij}}{\sum\limits_{i,j} [E_i (f_if_j)^{q-1} C_{ij} - \mu (f_if_j)^{q-1} C_{ij}]}
\end{equation}
But $\sum\limits_{i,j} C_{ij} = 0$ and also $(f_if_j)^{q-1} = (f_jf_i)^{q-1}$. So term with $\mu$ in the denominator becomes zero and hence we get
\begin{equation}
    \frac{\partial \epsilon}{\partial s}\biggr\vert_{n'} = T
\end{equation}
which proves that the relation Eq.~(\ref{eq:rel1}) is satisfied for unified distribution.

 \subsubsection{Relation 2}
Second thermodynamic relation is given as:
 \begin{equation}
    \frac{\partial \epsilon}{\partial n'}\biggr\vert_s = \mu
\end{equation}
\begin{equation}
     \frac{\partial E}{\partial N}\biggr\vert_s = \frac{\frac{\partial E}{\partial T}dT+{\frac{\partial E}{\partial \mu}d \mu}}{\frac{\partial N}{\partial T}dT+{\frac{\partial N}{\partial \mu}d \mu}}
\end{equation}
\begin{equation}
     \frac{\partial E}{\partial N}\biggr\vert_s = \frac{\frac{\partial E}{\partial T}+{\frac{\partial E}{\partial \mu}\frac{d \mu}{dT} }}{\frac{\partial N}{\partial T}+{\frac{\partial N}{\partial \mu}\frac{d \mu}{dT}}}
\end{equation}
Here $s$ is constant so 
\begin{equation}
ds = \frac{\partial s}{\partial T}dT +  \frac{\partial s}{\partial \mu}d\mu = 0
\end{equation}
\begin{equation}
    \frac{d \mu}{d T} = -\frac{\frac{\partial s}{\partial T}}{\frac{\partial s}{\partial \mu}}
\end{equation}
\begin{equation}
    \frac{\partial E}{\partial T}+{\frac{\partial E}{\partial \mu}\frac{d \mu}{dT} } =\frac{\sum\limits_{i,j} \left( \frac{qE_if_i^{q-1}}{\left(B f_{E_j}\right)^{\frac{1}{q} - 1}} - qE_i (f_if_j)^{q-1}\right)C_{ij}}{\sum\limits_{j}\biggr( \frac{1}{\left(B f_{E_j}\right)^{\frac{1}{q} - 1}}\frac{\partial f_j}{\partial \mu} -  f_j^{q-1}\frac{\partial f_j}{\partial \mu}\biggr)}
\end{equation}
 Similarly for  N part we get
    \begin{equation}
       \frac{\partial N}{\partial T}+{\frac{\partial N}{\partial \mu}\frac{d \mu}{dT}} =  \frac{\sum\limits_{i,j} \left( \frac{qf_i^{q-1}}{\left(B f_{E_j}\right)^{\frac{1}{q} - 1}} - q (f_if_j)^{q-1}\right)C_{ij}}{\sum\limits_{j}\biggr(\frac{1}{\left(B f_{E_j}\right)^{\frac{1}{q} - 1}}\frac{\partial f_j}{\partial \mu} -  f_j^{q-1}\frac{\partial f_j}{\partial \mu}\biggr)}
    \end{equation}
    On dividing above equations we get
     \begin{equation}
         \frac{\frac{\partial E}{\partial T}+{\frac{\partial E}{\partial \mu}\frac{d \mu}{dT} }}{\frac{\partial N}{\partial T}+{\frac{\partial N}{\partial \mu}\frac{d \mu}{dT}}} = \frac{\sum\limits_{i,j} \left( \frac{qE_if_i^{q-1}}{\left(B f_{E_j}\right)^{\frac{1}{q} - 1}} - qE_i (f_if_j)^{q-1}\right)C_{ij}}{\sum\limits_{i,j} \left( \frac{qf_i^{q-1}}{\left(B f_{E_j}\right)^{\frac{1}{q} - 1}} - q (f_if_j)^{q-1}\right)C_{ij}}
    \end{equation}
    Equation above reduces to
    \begin{equation}
          \frac{\partial \epsilon}{\partial n'}\biggr\vert_s = \mu
    \end{equation}
 \subsubsection{Relation 3}
Now considering the third relation:
\begin{equation}
 \frac{\partial P}{\partial \mu}\biggr\vert_T = n'
\end{equation}
where $n'$ is the number density. From laws of thermodynamics we get
\begin{equation}
P = \frac{-E + TS + \mu N}{V}
\end{equation}
\begin{equation}
 \frac{\partial P}{\partial \mu}\biggr\vert_T  = \frac{1}{V}\left[ \frac{-\partial E}{\partial \mu} + T\frac{\partial S}{\partial \mu} + N + \mu \frac{\partial N}{\partial\mu}\right] 
\end{equation}
\begin{equation}
\begin{aligned}
\frac{\partial P}{\partial \mu}\biggr\vert_T = \frac{1}{V} \sum\limits_{i} \biggr[ f_i^q -  \frac{T}{q-1} & \left\lbrace 1 + (q-1)\frac{(E_i -\mu)}{T}\right\rbrace \frac{\partial f_i^q}{\partial \mu} \\
& + \frac{Tq}{q-1} \frac{1}{\left(B f_{E_i}\right)^{\frac{1}{q}-1} } \frac{\partial f_i}{\partial\mu} \biggr]
\end{aligned}
\end{equation}
\begin{equation}
f_i = \left(B f_{E_i}\right)^{\frac{1}{q}} f_{Ta_i} 
\end{equation}
\begin{equation}
\frac{\partial f_i}{\partial \mu} = \frac{\left(B f_{E_i}\right)^{\frac{1}{q}}}{T} f_{Ta_i}^q
\end{equation}
\begin{equation}
\frac{\partial f_i^q}{\partial \mu} =B q \frac{f_{E_i}}{T} f_{Ta_i}^{2q-1}
\end{equation}
On substitution we will get,
\begin{equation}
\frac{\partial P}{\partial \mu}\biggr\vert_T  = \frac{1}{V} \sum\limits_{i} f_i^q = \frac{N}{V}
\end{equation}
\begin{equation}
 \frac{\partial P}{\partial \mu}\biggr\vert_T = n'
\end{equation}
\subsubsection{Relation 4}
Last equation which relates the derivative of pressure to the entropy density is given as:
\begin{equation}
   \frac{\partial P}{\partial T}\biggr\vert_\mu = s
\end{equation}
\begin{equation}
    \frac{\partial P}{\partial T} = \frac{1}{V} \left[-\frac{\partial E}{\partial T} + S + T\frac{\partial S}{\partial T} + \mu \frac{\partial N}{\partial T}\right]
\end{equation}
So in order to prove above relation we have to basically prove that
\begin{equation}\label{eq:rel4_criteria}
    -\frac{\partial E}{\partial T} + T \frac{\partial S}{\partial T} + \mu \frac{\partial N}{\partial T} = 0
\end{equation}

Solving the equation above we get
\begin{equation}
\begin{aligned}
=   \sum\limits_{i}  \biggr[- qf_i^{q-1}E_i\frac{\partial f_i}{\partial T} &+ \frac{Tq}{(q-1)\left(B f_E\right)^{\frac{1}{q} - 1}}\frac{\partial f_i}{\partial T}\\
& - \frac{Tqf_i^{q-1}}{q-1} \frac{\partial f_i}{\partial T} + \mu qf_i^{q-1}\frac{\partial f_i}{\partial T}\biggr] 
\end{aligned}
\end{equation}
On solving above equation we get zero which satisfies Eq.~(\ref{eq:rel4_criteria}). Hence,
\begin{equation}
   \frac{\partial P}{\partial T}\biggr\vert_\mu = s
\end{equation}
This proves that the unified distribution along with the form of entropy given in Eq.~(\ref{eq:pear_entropy}) is thermodynamically consistent. As shown above, the pass of thermal test makes the unified distribution thermodynamically relevant equations for a generalization by fulfilling all required criteria.  

As it has been discussed earlier, the relevance of Tsallis distribution is only limited to low-\pt region because of the dominance of hard processes in high-\pte. Unified distribution resolves this issue by generalizing the Tsallis distribution to include hard processes as well. In the next section, we have performed a comparative study of different distribution to prove that the unified distribution is indeed the best fit to explain transverse momentum spectra.

\section{\label{sec:result}RESULTS}

In order to test the applicability of unified distribution, we have performed a detailed analysis of the invariant yield of $\pi^+$ over different energy ranges.  The results shown below present the main body of validation obtained in this work where, a comparison between BG, Tsallis and unified statistical approaches in describing the transverse momentum spectra is demonstrated. It shows the degree of agreement between measured data and the results attainable by the approaches based on statistical thermodynamics. In the plots, symbols represent the experimentally measured data of transverse momentum and solid lines represent the results fitted by BG, Tsallis and unified distribution functions. The ROOT \cite{root} data analysis framework has been used along with MINUIT \cite{James:1975dr} class for fitting.

The analysis was done for the transverse momentum data of $\pi^+$ particles produced in $Au-Au$ and $Pb-Pb$ collisions and the collision energies we selected for study included $19.6$ GeV \cite{Adamczyk:2017iwn}, $27$ GeV\cite{Adamczyk:2017iwn}, $39.0$ GeV \cite{Adamczyk:2017iwn}, $130$ GeV \cite{Adcox:2001mf}, $200.0$ GeV \cite{Adler:2003cb} and $2.76$ TeV \cite{Abelev:2013vea}. 

The goodness of unified function approach over other approaches is determined by analysing the \textit{chi-square} values of each fit. We use the \textit{chi-square} goodness of fit test to determine how the observed value is different from the expected value and to compare the observed sample distribution with the expected probability distribution. A table including the Chi-square values of Boltzmann, Tsallis and unified functions fitted to \pte-spectra at several energies is given in Table \ref{tab:table3}.

 \begin{table}
\caption{\label{tab:table3}The $\chi^{2}/NDF$ values of transverse momentum data of $\pi^+$ particles fitted to Boltzmann, Tsallis and unified functions at various collision energies.}
\begin{tabular*}{\columnwidth}{@{\extracolsep{\fill}}llll@{}}
\hline
\multirow{2}{*}{$\sqrt{s_{NN}}$ (GeV)}
& \multicolumn{3}{c}{$\chi^2/NDF$}  \\
\cline{2-4}
&Boltzmann & Tsallis & Unified

\\ 
\hline
19.6&9.769&0.392&0.052\\
 27 & 9.934 & 0.316 & 0.040 \\
 39 & 10.299 & 0.275 & 0.003  \\
 130 & 45.747 & 5.118 & 1.912  \\
 200  & 337.676  & 14.567 & 1.798 \\
 2760 & 23.980 & 2.314 & 0.064 \\ 
\hline
\end{tabular*}
  
\end{table}
%
%
\begin{figure}[h!]
  \includegraphics[height=2in,width=3in]{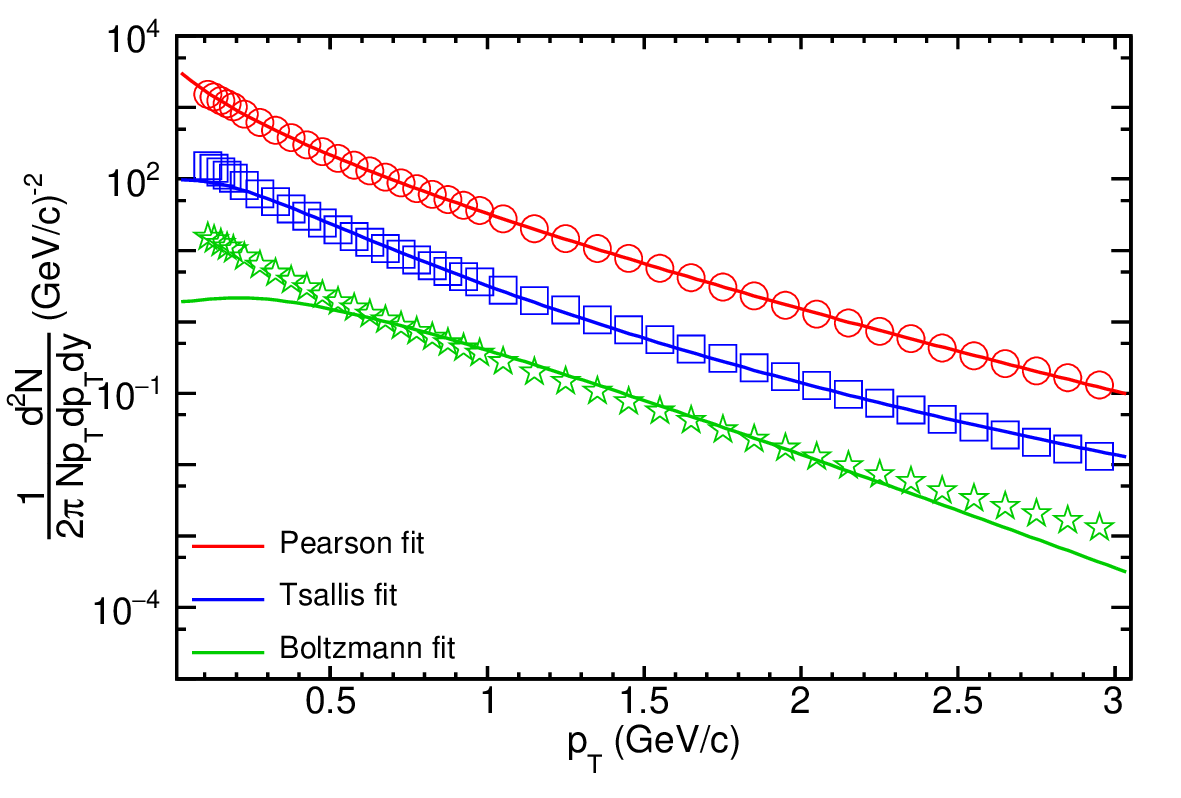}
  \caption{The transverse momentum spectra of $\pi^+$ particles produced in most central 2.76 TeV $Pb-Pb$ collision measured by the ALICE experiment \cite{Abelev:2013vea} fitted with Boltzmann Eq.~(\ref{eqn:BG}), Tsallis Eq.~(\ref{TSALLIS}) and unified distribution function Eq.~(\ref{PEARSON}). Data points are scaled for better visibility.}\label{fig:2760}
\end{figure}

\begin{figure}[h!]
  \includegraphics[height=2in,width=3in]{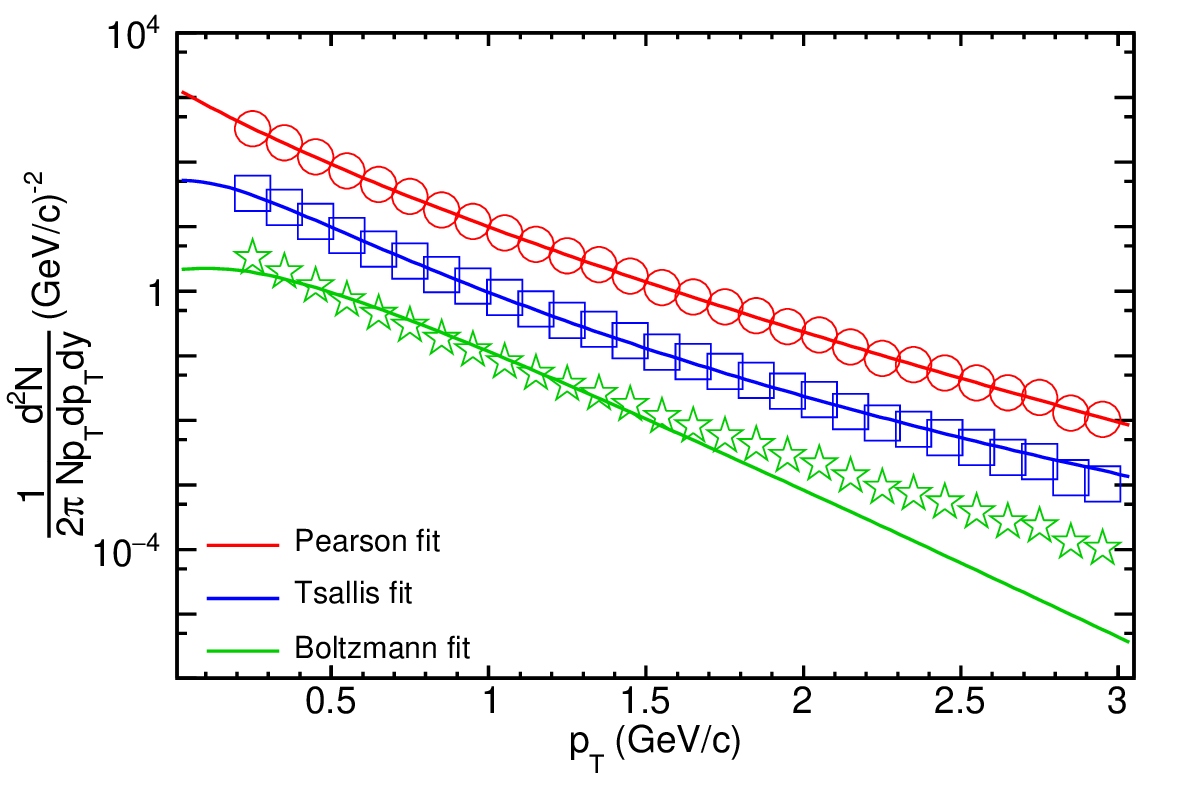}
  \caption{The transverse momentum data of $\pi^+$ particles produced in most central 200 GeV $Au-Au$ collision measured by the PHENIX experiment \cite{Adler:2003cb} fitted with Boltzmann Eq.~(\ref{eqn:BG}), Tsallis Eq.~(\ref{TSALLIS}) and unified distribution function Eq.~(\ref{PEARSON}). Data points are scaled for better visibility.}\label{fig:200}
\end{figure}
%

\begin{figure}[h!]
  \includegraphics[height=2in,width=3in]{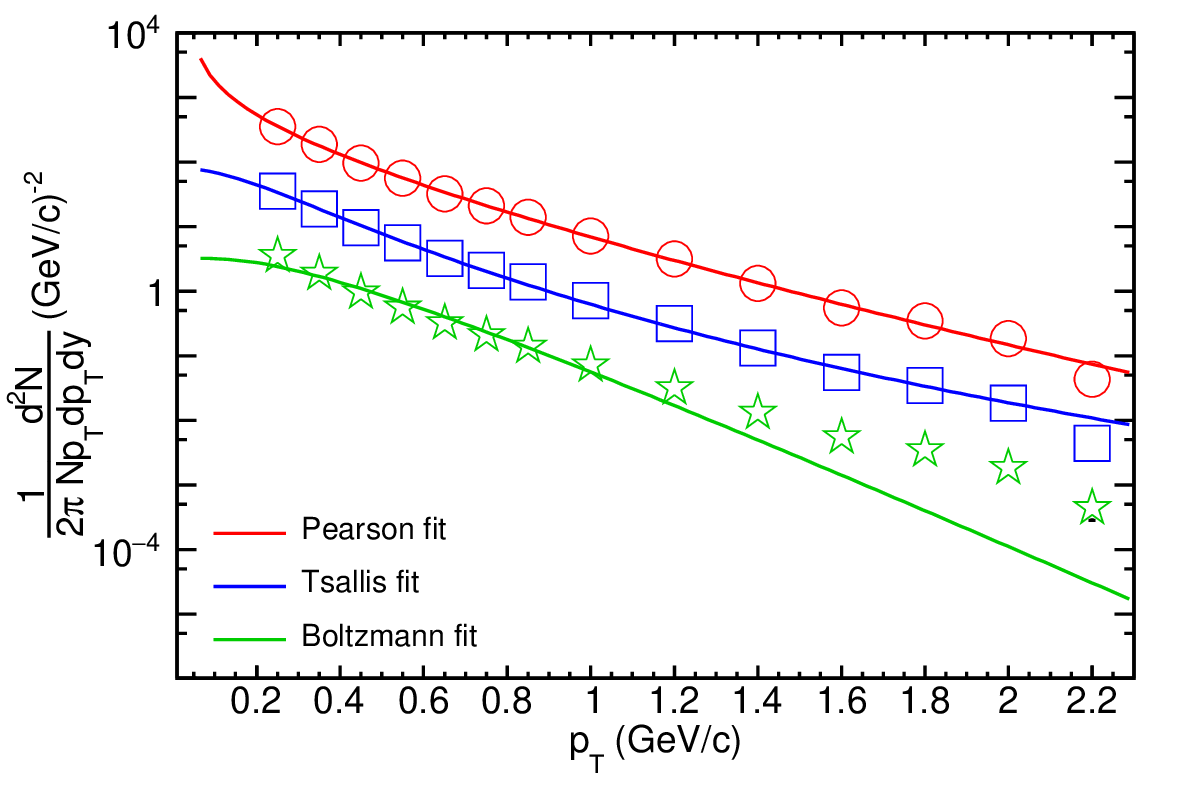}
  \caption{The transverse momentum data of $\pi^+$ particles produced in most central 130 GeV $Au-Au$ collision measured by the PHENIX experiment \cite{Adcox:2001mf} fitted with Boltzmann Eq.~(\ref{eqn:BG}), Tsallis Eq.~(\ref{TSALLIS}) and unified distribution function Eq.~(\ref{PEARSON}). Data points are scaled for better visibility.}\label{fig:130}
\end{figure}
\begin{figure}[h!]
  \includegraphics[height=2in,width=3in]{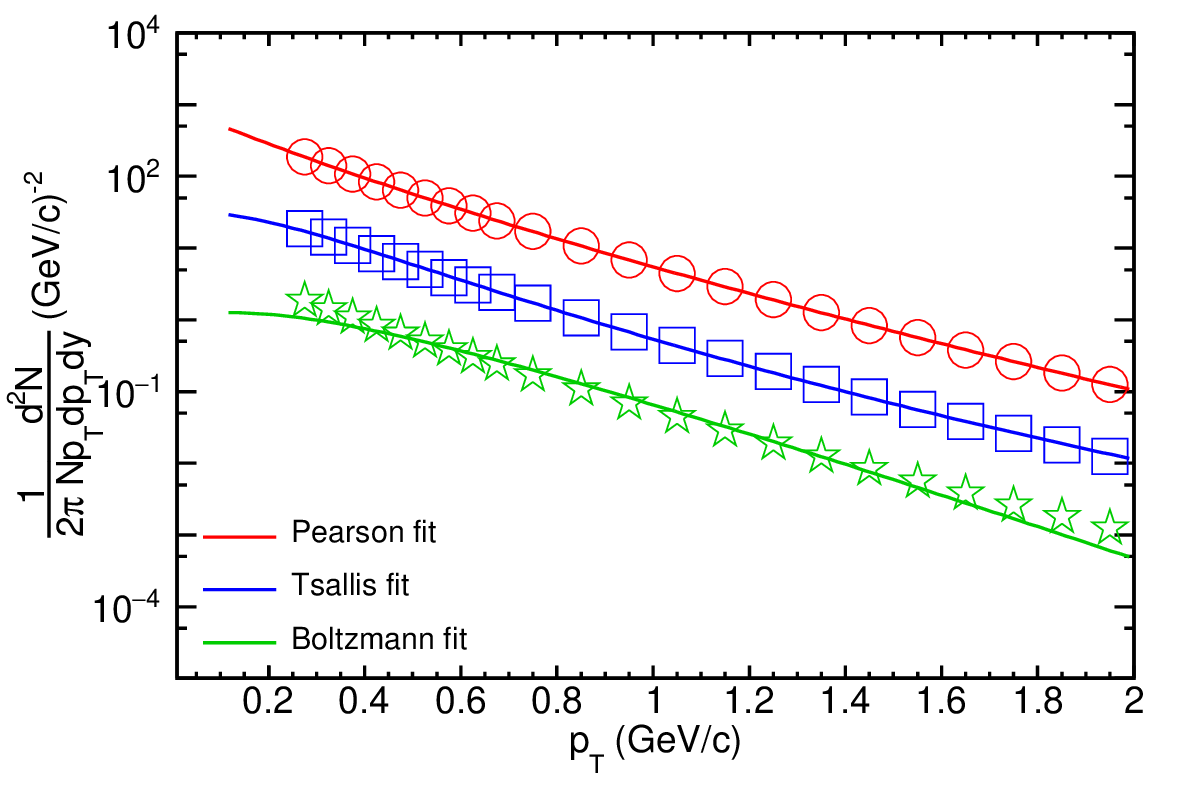}
 \caption{The transverse momentum data of $\pi^+$ particles produced in most central 39 GeV $Au-Au$ collision measured by the STAR experiment \cite{Adamczyk:2017iwn} fitted with Boltzmann Eq.~(\ref{eqn:BG}), Tsallis Eq.~(\ref{TSALLIS}) and unified distribution function Eq.~(\ref{PEARSON}). Data points are scaled for better visibility.}\label{FIG1960}
\end{figure}
\begin{figure}[h!]
  \includegraphics[height=2in,width=3in]{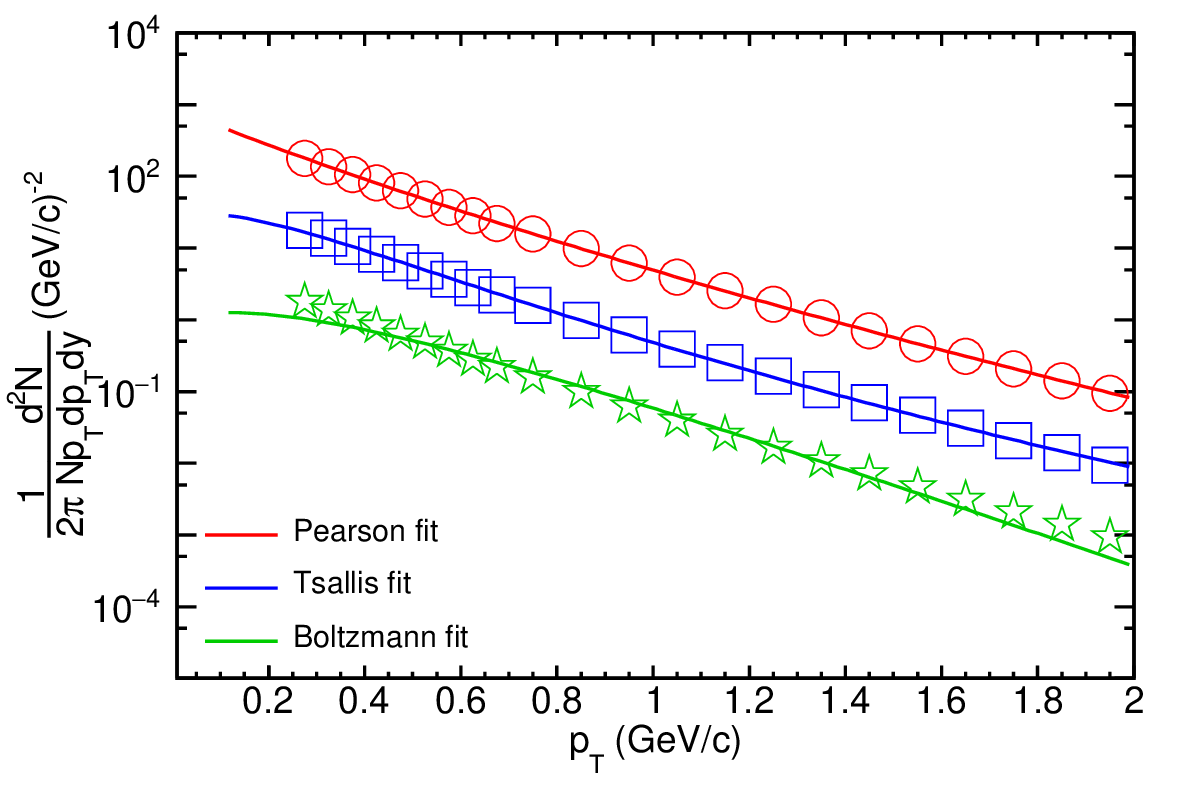}
  \caption{The transverse momentum data of $\pi^+$ particles produced in most central 27 GeV $Au-Au$ collision measured by the STAR experiment \cite{Adamczyk:2017iwn} fitted with Boltzmann Eq.~(\ref{eqn:BG}), Tsallis Eq.~(\ref{TSALLIS}) and unified distribution function Eq.~(\ref{PEARSON}). Data points are scaled for better visibility.}\label{FIG1150}
\end{figure}
\begin{figure}[h!]
  \includegraphics[height=2in,width=3in]{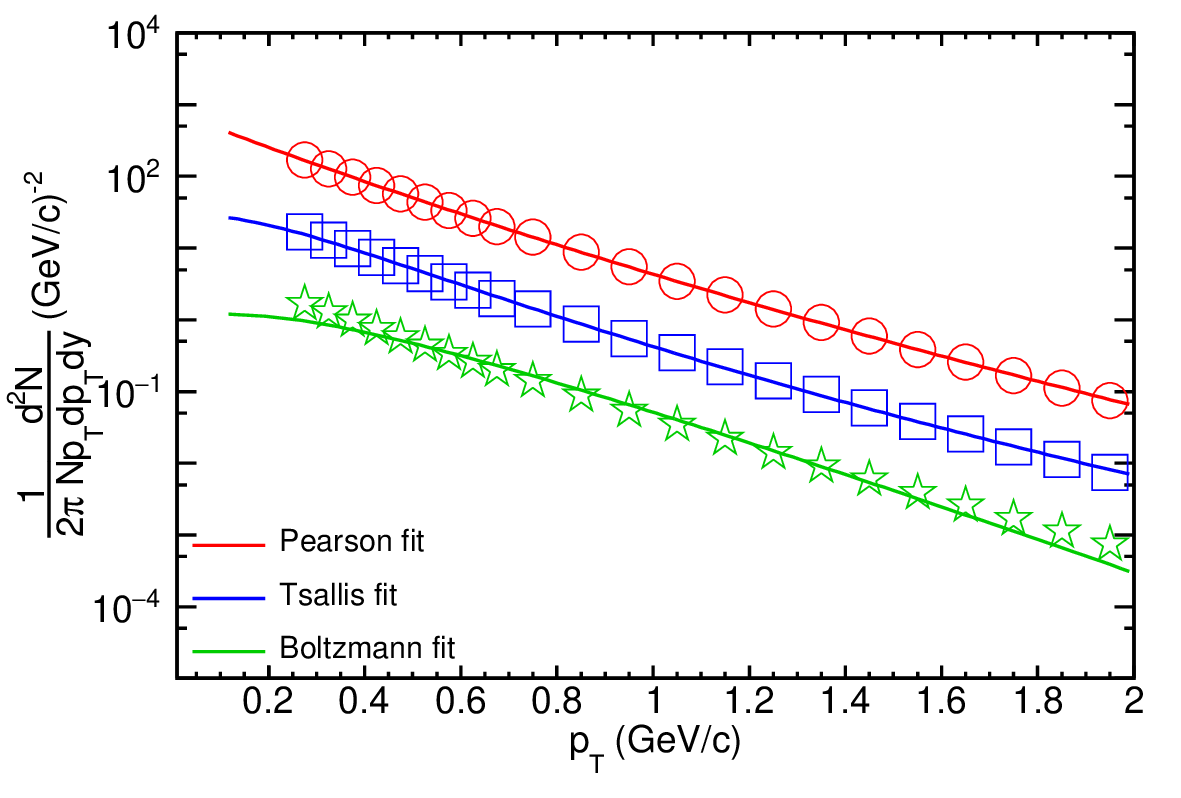}
  \caption{The transverse momentum data of $\pi^+$ particles produced in most central 19.6 GeV $Au-Au$ collision measured by the STAR experiment \cite{Adamczyk:2017iwn} fitted with Boltzmann Eq.~(\ref{eqn:BG}), Tsallis Eq.~(\ref{TSALLIS}) and unified distribution function Eq.~(\ref{PEARSON}). Data points are scaled for better visibility.}\label{FIG770}
\end{figure} 

 It is clearly evident from the fits in figures  \ref{fig:2760}, \ref{fig:200}, \ref{fig:130}, \ref{FIG1960}, \ref{FIG1150} $\&$ \ref{FIG770} that the unified fits are better compared to Boltzmann and Tsallis fits. This can be confirmed from the Table \ref{tab:table3}, where the $\chi^2/NDF$ values of all the fits are displayed. From table \ref{tab:table3} we observe that the $\chi^2/NDF$ values are minimum for unified fits at all the energies. As we expect, these are large for Boltzmann fits and that of Tsallis fits are intermediate. 
 
Boltzmann distribution is parameterized by only one parameter, which is the temperature (T). Tsallis framework includes another parameter called q-parameter apart from the temperature, which gives the extent of non-extensivity in the distribution. 
The proposed approach using unified distribution comprises of two more free parameters in addition to the temperature and q-parameter. 
The relation of unified function parameters with the higher-order moments could be a reason for its success over other distribution, most of which depends primarily on mean and standard deviation as parameters. Further, we observe that the Tsallis distribution deviates from data at high-\pt region, which form the tail part of the distribution. And the tail part of a distribution is more sensitive to the higher-order moments. This could be a statistical reason for the success of unified distribution, especially in high-\pt regime. 

The presence of quenching effect beyond certain \pt value in heavy-ion collision limit the application of the statistical thermal models, however, the absence of such effect in \pp collision makes it an ideal choice to test whether the developed formalism covers a broader range of \pte. Hence, to check the applicability of unified statistical framework over a broad \pt range, we have considered the \pp collision data, with \pt upto few hundred GeV/c, and the yield spanning over several order of magnitude. For this analysis, we have used the data of transverse momentum spectra of charged hadron produced in \pp~collision at four different energies ($\sqrt{s_{NN}} = 900$ GeV \cite{Chatrchyan:2011av}, $ 2.76$ TeV \cite{CMS:2012aa}, $5.02$ TeV \cite{Khachatryan:2016odn} and $7$ TeV  \cite{Chatrchyan:2011av}) measured by CMS experiment  over wide \pt range upto 400 GeV/c. Further, we have performed an analysis on the recently released high multiplicity \pp~collision data at $7$ TeV measured by ALICE experiment in different V0M event multiplicity classes \cite{Acharya:2018orn} with the multiplicities corresponding to each class provided in table \ref{mult_class}. The pseudorapidity ranges of data at $0.9$ TeV \&  $7$ TeV is $|\eta|<2.4$ \cite{Chatrchyan:2011av}, $2.76$ TeV \cite{CMS:2012aa} \& $5.02$ TeV \cite{Khachatryan:2016odn} is $|\eta|<1$. At the same time, the corresponding range for the multiplicity class divided data measured by ALICE experiment at $7$ TeV \cite{Acharya:2018orn} is $|\eta|<0.5$.

\begin{table}
\centering
\caption{VZERO multiplicity classes and the corresponding multiplicity values $\langle dN_{ch}/d\eta \rangle$}
\label{mult_class}
\begin{tabular*}{\columnwidth}{@{\extracolsep{\fill}}ll@{}}
\hline
\multicolumn{1}{@{}l}{Multiplicity class} &$7$ TeV \pp~collision \\
\hline
V0M \RomanNumeralCaps{1}      & $21.3 \pm 0.6$  \\
V0M \RomanNumeralCaps{2}      & $16.5 \pm 0.5$ \\
V0M \RomanNumeralCaps{3}      & $13.5 \pm 0.4$  \\
V0M \RomanNumeralCaps{4}      & $11.5 \pm 0.3$ \\
V0M \RomanNumeralCaps{5}      & $10.1 \pm 0.3$  \\
V0M \RomanNumeralCaps{6}      & $8.45 \pm 0.25$ \\
V0M \RomanNumeralCaps{7}      & $6.72 \pm 0.21$  \\
V0M \RomanNumeralCaps{8}      & $5.4 \pm 0.17$  \\
V0M \RomanNumeralCaps{9}      & $3.9 \pm 0.14$  \\
V0M \RomanNumeralCaps{10}     & $2.26 \pm 0.12$  \\
\hline
\hline
\end{tabular*}
\end{table}

\begin{figure}
	\centering
		\includegraphics[scale=.42]{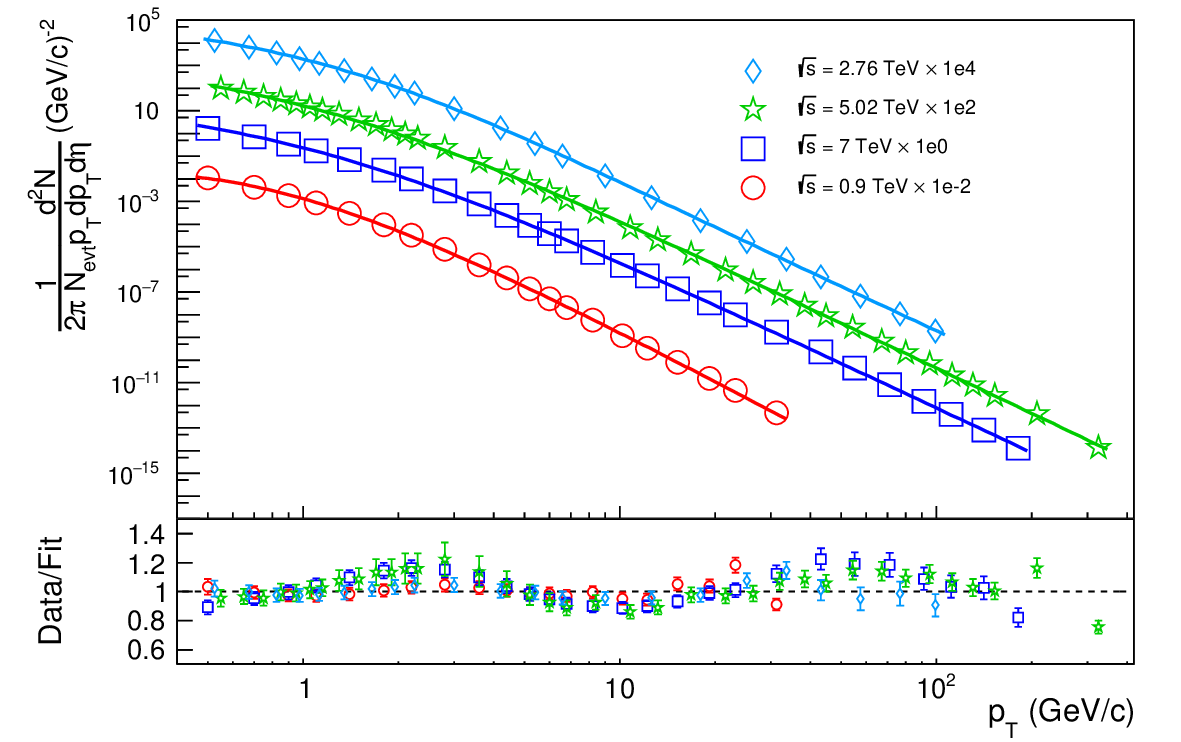}
	\caption{(color online) Top plot: The transverse momentum data of charged hadrons produced in \pp~collision at $0.9$ TeV \cite{Chatrchyan:2011av}, $ 2.76$ TeV \cite{CMS:2012aa}, $5.02$ TeV \cite{Khachatryan:2016odn} and $7$ TeV  \cite{Chatrchyan:2011av} measured by the CMS experiment fitted with unified distribution Eq.~(\ref{pearson_final}). Bottom plot: Ratio of the experimental data to the corresponding value obtained from the fit function. }
\label{fig:pearson_4energy}
\end{figure}
\begin{figure}
	\centering
		\includegraphics[scale=.42]{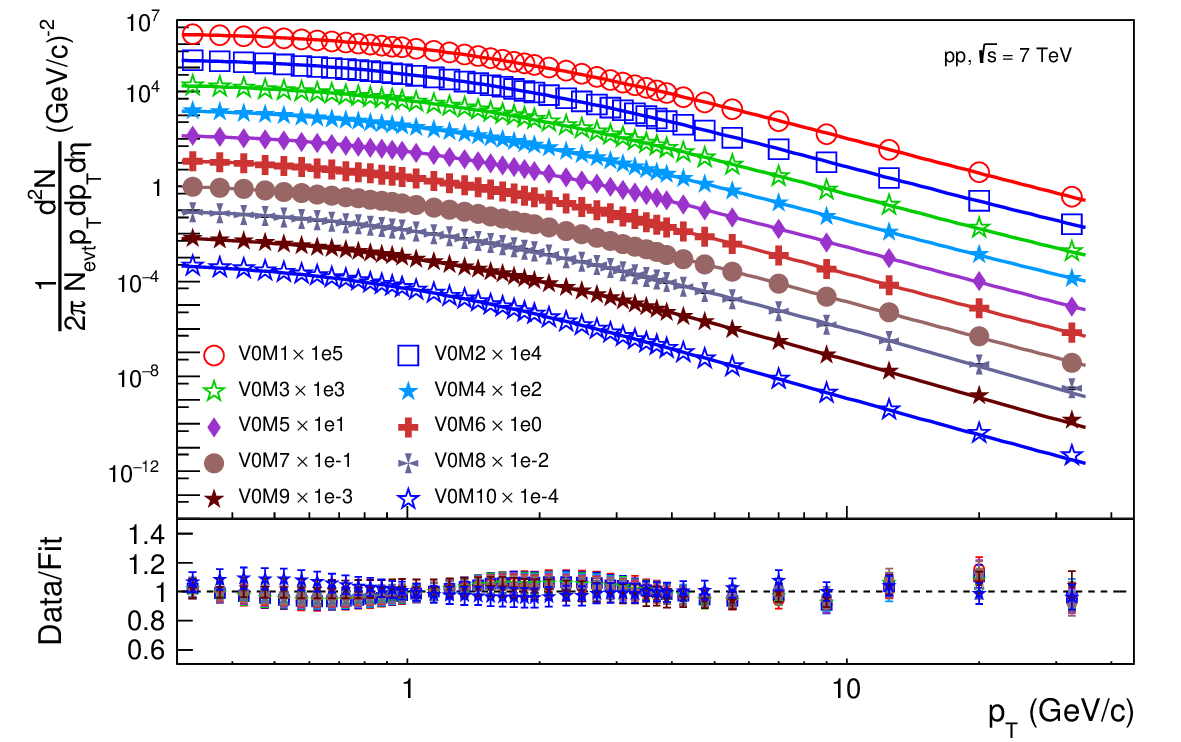}
	\caption{(color online) Top plot: The transverse momentum data of charged hadrons divided into multiplicity classes produced in \pp~collision at $7$ TeV \cite{Acharya:2018orn} measured by the ALICE experiment fitted with unified distribution Eq.~(\ref{pearson_final}). Bottom plot: Ratio of the experimental data to the corresponding value obtained from the fit function. }
\label{fig:pearson_7TeV} 
\end{figure}

Figure \ref{fig:pearson_4energy} represents the unified function fit to \pt spectra for four different energies with \pt range upto $400$ GeV. The corresponding fit to $7$ TeV data divided into separate multiplicity classes is given in Fig.~\ref{fig:pearson_7TeV}. From the plot of the ratio of experimental data to the fit function, we observe a log-periodic oscillation over a broad range of transverse momenta for the unified statistical framework. This form of oscillation has been discussed for Tsallis distribution in Ref.~\cite{Azmi:2015xqa,Wilk:2014sza,Wilk:2014bia}. Further, the oscillation observed in the $7$ TeV ALICE experiment data shows an interesting pattern over different multiplicity classes. Here we observe a clear reversal in the oscillation pattern as we go from the ALICE multiplicity class V0M 1 to V0M 10. This strange behaviour in data over fit needs to be further explored, and it has the potential to give interesting physics information.

In table \ref{fit_val_energy} and \ref{fit_val}, we have provided the fitted value of temperature and q parameters that appear in the unified function Eq.~(\ref{pearson_final}) and the $\chi^2/NDF$ values, which represent the goodness of the fit. Low $\chi^2/NDF$ values in the tables suggest a good agreement between the experimental data and the unified distribution function. 
Ratio plot of different energies for unified function fit (Fig.~\ref{fig:pearson_4energy} and Fig.~\ref{fig:pearson_7TeV}) also compliments the observation of goodness of fit to the experimental data.

\begin{table}
\centering
\caption{Best fit value of the parameters $T$ (GeV) $\&~q$ and the $\chi^2/NDF$ value obtained by fitting the charged hadron spectra produced in $pp$ collision  at $0.9$ TeV \cite{Chatrchyan:2011av}, $ 2.76$ TeV \cite{CMS:2012aa}, $5.02$ TeV \cite{Khachatryan:2016odn} and $7$ TeV  \cite{Chatrchyan:2011av} measured by the CMS experiment with the unified distribution function Eq.~(\ref{pearson_final}).}
\label{fit_val_energy}
\begin{tabular*}{\columnwidth}{@{\extracolsep{\fill}}llll@{}}
\hline
\multicolumn{1}{@{}l}{Energy} &T & q & $\chi^2/NDF$ \\
\hline
\multirow{2}{*}{0.9 TeV}      & 0.078 & 1.032 &
\multirow{2}{*}{1.790} \\
&$\pm0.009$ &$\pm 0.003$ & \\
\hline
\multirow{2}{*}{2.76 TeV}      & 0.132 & 1.070 & 
\multirow{2}{*}{0.996} \\
&$\pm0.006$ &$\pm 0.002$ & \\
\hline
\multirow{2}{*}{5.02 TeV}      & 0.146 & 1.122  &
\multirow{2}{*}{3.119} \\
&$\pm0.007$ &$\pm 0.001$ &  \\
\hline
\multirow{2}{*}{7 TeV}      & 0.125 & 1.147  &
\multirow{2}{*}{4.559} \\
&$\pm0.001$ &$\pm 0.001$ &  \\
\hline
\hline
\end{tabular*}
\end{table}

\begin{table}
\centering
\caption{Best fit value of the parameters $T$ (GeV), $q$ and the $\chi^2/NDF$ value obtained by fitting the multiplicity class divided charged hadron spectra produced in $pp$ collision  at $7$ TeV measured by the ALICE experiment \cite{Acharya:2018orn} with the unified distribution function Eq.~(\ref{pearson_final}).}
\label{fit_val}
\begin{tabular*}{\columnwidth}{@{\extracolsep{\fill}}llll@{}}
\hline
\multicolumn{1}{@{}l}{Mult.~class} &T & q & $\chi^2/NDF$ \\
\hline
\multirow{2}{*}{V0M \RomanNumeralCaps{1}}      & 0.221 & 1.146  & \multirow{2}{*}{0.996} \\
&$\pm0.011$ &$\pm 0.004$  & \\
\hline
\multirow{2}{*}{V0M \RomanNumeralCaps{2}}      & 0.211& 1.145&  \multirow{2}{*}{0.787} \\
&$\pm0.010$ &$\pm0.004$  &\\
\hline
\multirow{2}{*}{V0M \RomanNumeralCaps{3}}      &0.202& 1.142  &\multirow{2}{*}{0.639}   \\
& $\pm0.011$ & $\pm0.005$ & \\
\hline
\multirow{2}{*}{V0M \RomanNumeralCaps{4}}      & 0.194 & 1.132  & \multirow{2}{*}{0.518} \\
& $\pm0.010$ & $\pm0.005$  & \\
\hline
\multirow{2}{*}{V0M \RomanNumeralCaps{5}}      & 0.190 & 1.136  & \multirow{2}{*}{0.518}   \\
& $\pm0.017$ & $\pm 0.009$ & \\
\hline
\multirow{2}{*}{V0M \RomanNumeralCaps{6}}      & 0.182 & 1.129  & \multirow{2}{*}{0.321}  \\
& $\pm0.017$ & $\pm0.009$  & \\
\hline
\multirow{2}{*}{V0M \RomanNumeralCaps{7}}      & 0.166 & 1.114  & \multirow{2}{*}{0.337}  \\
& $\pm0.003$ & $\pm0.001$  & \\
\hline
\multirow{2}{*}{V0M \RomanNumeralCaps{8}}      &  0.167 & 1.121 &  \multirow{2}{*}{0.107} \\
&$\pm 0.005 $& $\pm0.002$ &  \\
\hline
\multirow{2}{*}{V0M \RomanNumeralCaps{9}}      & 0.156 & 1.135  & \multirow{2}{*}{0.377}  \\
&$\pm0.005$ & $\pm0.003$ &\\
\hline
\multirow{2}{*}{V0M \RomanNumeralCaps{10}}     & 0.126 & 1.077  & \multirow{2}{*}{0.726}  \\
&$\pm0.005$ & $\pm0.002$ & \\
\hline
\hline
\end{tabular*}
\end{table}

  \section{Discussion}
  Unified statistical framework is a more generalized approach to statistically explain the system created in ultra high energy collisions. What distinguishes unified formalism from a simple polynomial fit is the richness of physics it incorporates. It is a thermodynamically consistent formalism following the laws of thermodynamics. The non-extensivity properties of the unified statistics evolved similarly to that of Tsallis statistics, however, one of the important distinction between the two is the presence of additional parameter $p_0$ and $n$ in the former case whose connection with the physics observable needs to be explored to get the ultimate benefit of this formalism. One such effort has been made in Ref.~\cite{Jena:2020wno}, where the unified function parameter is shown to nicely describe the second-order flow parameter $v_2$. A linear relationship has been established between the $v_2$ and the unified function parameter $n$ using the charged hadron spectra produced in $Pb-Pb$ collision at $2.76$ TeV.
  
 The presence of more free parameters in unified formalism gives an extra advantage in terms of obtaining better fit results as compared to Tsallis statistics. This can be verified from the result obtained in the previous section, where we observe an improvement in the quality of fit using unified formalism than the Tsallis statistics. This improvement can also be attributed to the unique approach where the physics of soft and hard processes are considered in a unified manner. We have also tested the applicability of unified formalism in small collision system and we have shown that the unified formalism provide better fit to the \pp collision data upto few hundreds of $GeV/c$ and over different energies and multiplicities.
  
 One can also argue that a higher-order polynomial may do a better fitting; however, the richness of physics that unified formalism incorporates make it a better choice being thermodynamically consistent.  Also, unified formalism is backward compatible with Tsallis distribution under the limiting condition of the parameters. From the statistical perspective, the relation of unified distribution with higher-order moments such as skewness and kurtosis also makes it a suitable choice to fit a skewed dataset. This further strengthens the need for exploration of the unified statistical framework in other areas of high energy physics.
\section{Conclusion}
In this work, we have presented a comprehensive study of the theoretical framework to analyze transverse momentum spectra. Further, we have provided the detailed mathematical description of the unified distribution which was first proposed in Ref.~\cite{Jena:2020wno} and is proved to describe both soft and hard scattering regions of particle spectra in a unified manner. We have also demonstrated the applicability of unified distribution in the study the \pte-spectra of different particle produced in heavy-ion collision as well as the \pp collision  over a broad energy range. Further, we have also explored the thermodynamical consistency of unified distribution and proved that this distribution consistently follows the laws of thermodynamics. 
 
 This formalism has very wide application in high energy physics. We can utilize this formalism to extract several thermodynamics quantities such as the isothermal compressibility, speed of sound \cite{Jain:2021qro}, chemical potential, specific heat of the system created in heavy-ion collision. This formalism can also be modified to study the pseudorapidity distribution \cite{Gupta:2021oxf} and the particle multiplicity.

 In conclusion, we would like to point out that we have proposed a generalization to Tsallis-like distribution function with the significant improvement in the goodness of fit to the spectra. We would also like to stress that although several theoretical and phenomenological works are proposed to study the particle production in high energy collision, more novel ideas are required to tap into the full potential of the data obtained in mega collider experiments.
 
 This paper provide a detailed theoretical description of the unified statistical framework and has the potential for wider applicability in other areas of physics. 
 
 \section{Acknowledgement}

  R. Gupta would like to acknowledge the financial support provided by CSIR through fellowship number 09/947 (0067) 2015-EMR-1. A. Menon would like to acknowledge that this work was done during her association at IISER Mohali as a part of her thesis work.


\end{document}